\documentclass[english,traditabstract]{aa}
\usepackage{mathptmx}
\usepackage[T1]{fontenc}
\usepackage[latin9]{inputenc}
\usepackage{graphicx}
\usepackage{amssymb}
\usepackage[authoryear]{natbib}

\usepackage{babel}

\begin{document}

\title{The effect of rotation on the stability of nuclear burning in accreting
neutron stars}

\author{L.~Keek\inst{1,2}\thanks{Present address: School of Physics and Astronomy, University of Minnesota, 116 Church Street S.E.,  Minneapolis, MN 55455, USA}\and N.~Langer\inst{1}\and J.~J.~M.~in~'t~Zand\inst{2}}

\institute{Astronomical Institute, Utrecht University, P.O. Box 80000, NL -
3508 TA Utrecht, the Netherlands \and SRON Netherlands Institute
for Space Research, Sorbonnelaan 2, NL - 3584 CA Utrecht, the Netherlands }

\offprints{laurens@physics.umn.edu}

\date{Received/Accepted}

\abstract{Hydrogen and/or helium accreted by a neutron star from a binary companion
may undergo thermonuclear fusion. At different mass accretion rates
different burning regimes are discerned. Theoretical models predict
helium fusion to proceed as a thermonuclear runaway for accretion
rates below the Eddington limit and as stable burning above this limit.
Observations, however, place the boundary close to 10\% of the Eddington
limit. We study the effect of rotationally induced transport processes
on the stability of helium burning. For the first time detailed calculations
of thin helium shell burning on neutron stars are performed using
a hydrodynamic stellar evolution code including rotation and rotationally
induced magnetic fields. We find that in most cases the instabilities
from the magnetic field provide the dominant contribution to the chemical
mixing, while Eddington-Sweet circulations become important at high
rotation rates. As helium is diffused to greater depths, the stability
of the burning is increased, such that the critical accretion rate
for stable helium burning is found to be lower. Combined with a higher
heat flux from the crust, as suggested by recent studies, turbulent
mixing could explain the observed critical accretion rate. Furthermore,
close to this boundary we find oscillatory burning, which previous
studies have linked to mHz QPOs. In models where we continuously lower
the heat flux from the crust, the period of the oscillations increases
by up to several tens of percents, similar to the observed frequency
drift, suggesting that this drift could be caused by the cooling of
deeper layers.
}

\maketitle

\section{Introduction}

Neutron stars in low-mass X-ray binaries (LMXBs) accrete matter from
a companion star through Roche-lobe overflow. Due to conservation
of angular momentum, the matter forms a disk around the neutron star,
where it radiates away a large fraction of the potential energy before
reaching the neutron star surface. The companion is typically a main
sequence star, donating matter with a large hydrogen abundance. For
a number of systems which likely have a white dwarf companion, hydrogen
may be deficient but helium not (\citealt{Nelson1986,Webbink,Zand2005}).

The accreted matter is thought to quickly spread over the neutron
star surface, forming a thin layer. When this layer is a few meters
thick, the temperature and density at the base reach the ignition
conditions for hydrogen and helium fusion. The thermonuclear burning
can proceed in a stable or an unstable manner, depending on the temperature
dependence of the burning and cooling rates, which depend on the conditions
in the thin burning shell. These conditions are largely set by the
mass accretion rate (\citealt{Fujimoto1981}; see also \citealt{Bildsten1998}).
Models predict that hydrogen burning via the CNO cycle is stable for
$\dot{M}\gtrsim10^{-2}\dot{M}_{\mathrm{Edd}}$ (assuming spherical
symmetry), while it is unstable for lower accretion rates (assuming
a neutron star mass of $1.4\,\mathrm{M_{\odot}}$ with a $10\,\mathrm{km}$
radius, a hydrogen mass fraction $X=0.7$ and a CNO mass fraction
$Z_{\mathrm{CNO}}=0.01$; $\dot{M}_{\mathrm{Edd}}$ is the critical
accretion rate where the accretion luminosity equals the Eddington
luminosity). For helium burning via the triple-alpha process, theory
places the critical accretion rate at the Eddington limit $\dot{M}\simeq\dot{M}_{\mathrm{Edd}}$
(\citealt{Fujimoto1981,Ayasli1982}), which suggests that stable helium
burning does not take place in practice. If the accreted matter is
hydrogen-deficient, the transition is expected to take place at an
even higher accretion rate due to the lack of heating by stable hydrogen
burning: $\dot{M}\simeq10\dot{M}_{\mathrm{Edd}}$ (\citealt{Bildsten1998}).

Nevertheless, there are observational indications that stable helium
burning takes place at lower accretion rates. Van~Paradijs et al.
\citeyearpar{Paradijs1988} observe from ten sources, most of which
are thought to accrete hydrogen-rich material, that the fraction of
hydrogen and helium that is burned in a stable manner increases strongly
with the mass accretion rate, while no bursts are observed at accretion
rates of $\dot{M}\gtrsim0.3\dot{M}_{\mathrm{Edd}}$. \citet{Cornelisse2003}
find for 4U~1820-30, which likely has a hydrogen-deficient mass donor,
that the burst rate is lower by a factor of $5$ for $\dot{M}\gtrsim0.07\dot{M}_{\mathrm{Edd}}$
than for lower mass accretion rates, while no bursts are observed
at $\dot{M}\gtrsim0.14\dot{M}_{\mathrm{Edd}}$. Therefore, the transition
of unstable to stable burning likely takes place within several tens
of percents of $0.1\dot{M}_{\mathrm{Edd}}$.

A possible solution for this discrepancy is the inclusion of mixing
processes. If hydrogen and/or helium are mixed to larger depths, where
temperature and density are higher, the conditions under which the
fuel is burned are different. An efficient source of mixing is formed
by rotationally induced hydrodynamic instabilities (\citealt{Fujimoto1993}).
\citet{Yoon2004} studied the effect of rotational mixing processes
on thin shell burning in accreting white dwarfs and found burning
to proceed more stably. \citet{Piro2007} calculated the effect of
turbulent mixing in neutron stars, including mixing due to rotationally
induced magnetic fields (\citealt{Spruit2002}). They find that the
magnetic fields are the main driver of the mixing. Their analysis
of helium accretion onto an iron layer shows that in the case of unstable
burning, rotationally induced mixing allows for lower recurrence times
for helium flashes. Furthermore, their models including mixing allow
for steady-state burning at accretion rates below $\dot{M}_{\mathrm{Edd}}$.
Analytically the dependence of $\dot{M}_{\mathrm{st}}$, the accretion
rate at the transition between stable and unstable burning, on the
rotation rate of the neutron star, $\Omega$, is derived as $\dot{M}_{\mathrm{st}}\propto\Omega^{0.62}$.
For neutron stars that rotate slower than 10\% of the Keplerian velocity
at the surface, \citet{Piro2007} find $0.25\,\dot{M}_{\mathrm{Edd}}\lesssim\dot{M}_{\mathrm{st}}\lesssim\dot{M}_{\mathrm{Edd}}$. 

\citet{Piro2007} created one-dimensional numerical models of the
neutron star envelope using analytical approximations of chemical
mixing and of helium burning. In this paper we study the effect of
mixing due to rotation and a rotationally induced magnetic field on
the stability of helium burning in the neutron star envelope in more
detail. We present the results of detailed numerical simulations,
where we use a one-dimensional multi-zone model of the outer layers
of the neutron star. We introduce both rotation and magnetic fields
in the model. Using a stellar-evolution code we investigate the stability
of the burning as a function of the mass accretion rate, the rotation
rate of the neutron star core and the heat flux emanating from the
crust.

\section{Method}

To model the outer layers of the neutron star we employ a stellar
evolution code that is a modified version of the code used by \citet{Yoon2004}
to model nuclear burning of helium shells in white dwarfs. It implicitly
solves the stellar structure equations on a one-dimensional grid consisting
of a series of mass shells. The effect of the centrifugal force on
the stellar structure is taken into account. Consequently, the mass
shells are isobars rather than spherical shells (cf. \citealt{Heger2000}).
The chemical evolution is modeled by an extensive nuclear network.
The effects of rotation on the chemical mixing and the transport of
angular momentum are calculated. The effects of a rotationally induced
magnetic field on the mixing of angular momentum and chemical species
are calculated as well (\citealt{Spruit2002}). Convection is taken
into account using the Ledoux criterion. Neutrino losses are calculated
using the results of \citet{Itoh1996}. We make use of the OPAL tables
(\citealt{Iglesias1996}) for determining opacities. Heating due to
viscous energy dissipation is not taken into account as it will prove
to be negligible with respect to the assumed flux from the crust or
the energy release from nuclear burning.

We consider the accretion of matter which is hydrogen-deficient, but
has a large (99\%) helium content. This allows us to study the stability
of helium shell burning without the complication of the nuclear processes
involving hydrogen. We will discuss the potential effect of the inclusion
of hydrogen on our results.

Relativistic effects are not considered during the simulation, since
they are small in the thin helium burning shell. The effect of general
relativity on observables, such as the luminosity and mass accretion
rate, can be taken into account with appropriate gravitational redshift
factors (e.g. \citealt{Woosley2004}), reducing the values for the
observables typically by several tens of percents. We do not apply
these corrections to our results, as we are primarily interested in
the relative effect of rotationally induced mixing on the stability
of shell burning.

As our models are one-dimensional, any asymmetries in the longitudinal
and latitudinal directions are not resolved. For example, burning
likely starts at one location in the envelope, after which it spreads
to other regions (\citealt{Bildsten1995,Spitkovsky2002,Cooper2007}).
This idea is supported by the observation of so-called burst oscillations
(\citealt{Strohmayer1996,str06}), which are periodic variations in
the light curve at the spin frequency of the neutron star that are
observed during some type-I bursts. Also, \citet{Bildsten1995} proposes
a regime where the burning spreads relatively slowly across the surface
on a similar time scale as the accretion of matter. We restrict ourselves
to the case of stable helium burning, where flame spreading is not
an issue.

\subsection{Grid}

The one-dimensional grid divides the star in shells with a certain
mass and is separated in two parts to facilitate matter accretion
while limiting numerical noise in the nuclear burning region. In the
inner Lagrangian part each shell is assigned an absolute amount of
mass, while in the outer so-called pseudo-Lagrangian part a shell
is assigned an amount of mass relative to the total mass in the model.
To minimize numerical diffusion, we choose these regions such that
the nuclear burning takes place in the Lagrangian part. After each
time step the grid is adjusted to ensure that any gradients in for
instance temperature, energy generation rate or chemical composition
are well resolved.

Apart from the different grid points our model has the notion of a
{}``core'', which represents the inner part of the star, that is
not modeled in detail and which sets the inner boundary conditions
of our grid. The core has a mass, radius, rotation rate and a luminosity.
These quantities are kept at fixed values. During consecutive time
steps angular momentum is exchanged between the grid and the core,
after which we return the angular momentum and velocity of the core
to the initial values, effectively preventing the core to spin up
or down. We can safely make this approximation, since the amount of
angular momentum that is accreted on the timescale of nuclear burning
is negligible with respect to the angular momentum of the core.

\subsection{Accretion}

Mass accretion is implemented by increasing the total amount of mass
and scaling the mass of the grid points in the pseudo-Lagrangian region.
The mass fractions of the isotopes at each of these grid points are
set to the weighted mean of the original composition of the grid point
and the assumed composition of the accreted matter, where we weigh
by, respectively, the original mass of the grid point and the accreted
mass that is added to the grid point.

The accreted matter originally carries angular momentum corresponding
to orbital motion at the Keplerian angular velocity at the surface
of $\Omega_{\mathrm{K}}=\sqrt{\mathrm{G}MR^{-3}}$, where $M$ and
$R$ are the mass and radius of the star, respectively. We assume,
however, that the angular momentum of the accreted material is quickly
shared with a number of the outermost grid points. This region coincides
with or is slightly smaller than the pseudo-Lagrangian region where
the mass is added. The distribution of the angular momentum is done
such that the specific angular momentum is constant in the outer region.
After the accreted angular momentum is added to the model, the mixing
of angular momentum resulting from the different viscous processes
is applied to the model. As in the treatment of the accretion of mass,
we make sure that the region where angular momentum is accreted lies
outside the region of interest, i.e. well above the helium burning
shell.

\subsection{Diffusivity and viscosity\label{sub:Diffusivity-and-viscosity}}

Due to the accretion of angular momentum, the outer layers of the
star spin faster than the core. Therefore a non-zero shear $q$ is
present\begin{equation}
q=\frac{\mathrm{d}\ln\Omega}{\mathrm{d}\ln r},\label{eq:shear}\end{equation}
where $r$ is the radial coordinate. The shear can give rise to hydrodynamical
instabilities. We consider the dynamical shear instability, the secular
shear and the Solberg-H\o iland instability as well as Eddington-Sweet
circulation (see, e.g., \citealt{Heger2000} for details of the implementation).
Furthermore, we consider the presence of a rotationally induced magnetic
field.

A magnetic field is employed following the prescription of the {}``Tayler-Spruit
dynamo'' presented by \citet{Spruit2002}. A small initial radial
magnetic field $B_{r}$ winds up due to the shear to create a much
stronger toroidal field $B_{\phi}$. The field growth is limited by
magnetohydrodynamic (MHD) instabilities. In this process $B_{r}$
is increased as well, such that once an equilibrium situation is reached,
the initial field strength $B_{r}$ is unimportant. Equilibrium is
reached on much shorter timescales than the mixing processes induced
by the magnetic field. Therefore, in our model we employ the equilibrium
values for $B_{r}$ and $B_{\phi}$.

Both the magnetic and non-magnetic hydrodynamic instabilities cause
turbulence which allows for chemical mixing and the transport of angular
momentum. Furthermore, angular momentum is transported as well due
to magnetic torques. We treat the chemical and angular momentum mixing
as diffusion processes. Transfer of angular momentum is governed by
the one-dimensional diffusion equation \begin{equation}
\frac{\mathrm{d}\Omega}{\mathrm{d}t}=\frac{1}{\rho r^{4}}\frac{\mathrm{d}}{\mathrm{d}r}\left(\rho r^{4}\nu\frac{\partial\Omega}{\partial r}\right),\label{eq:viscosity}\end{equation}
 with $\nu$ the viscosity and $\rho$ the density. Diffusion of the
mass fraction $Y$ of a given isotope, e.g. $^{4}$He, is calculated
via \begin{equation}
\frac{\mathrm{d}Y}{\mathrm{d}t}=\frac{1}{\rho r^{2}}\frac{\mathrm{d}}{\mathrm{d}r}\left(\rho r^{2}D\frac{\partial Y}{\partial r}\right),\label{eq:diffusivity}\end{equation}
 where $D$ is the diffusivity. $D$ and $\nu$ are measures for the
efficiency of the respective diffusion processes. $D$ is generally
smaller than $\nu$, since it requires more work to exchange material
than to exchange angular momentum. For the non-magnetic instabilities
we use $\nu=30D$ (\citealt{Chaboyer1992}). This is not the case
for the $D$ and $\nu$ that follow from the Tayler-Spruit mechanism,
where angular momentum is transported due to magnetic torques, such
that $D$ and $\nu$ differ in their dependencies on $q$ and $\Omega$.

The Tayler-Spruit formalism accounts for both compositional and a
temperature gradients, which provide a stabilizing stratification
against the magnetohydrodynamic instabilities that limit the growth
of the magnetic field. The effective radial viscosity $\nu_{\mathrm{re}}$
is described as a function of $\nu_{\mathrm{e0}}$ and $\nu_{\mathrm{e}1}$,
the viscosities when thermal diffusion either can be ignored or dominates,
respectively:\begin{equation}
\nu_{\mathrm{re}}=\frac{\nu_{\mathrm{e0}}\nu_{\mathrm{e1}}}{\nu_{\mathrm{e0}}+\nu_{\mathrm{e1}}}f(q),\label{eq:nu total}\end{equation}
where the factor $0<f(q)<1$ is introduced to ensure $\nu_{\mathrm{re}}$
vanishes smoothly below a certain minimum shear (\citealt{Spruit2002}).\begin{equation}
\begin{array}{l}
\nu_{\mathrm{e0}}=r^{2}\Omega q^{2}\left(\frac{\Omega}{N_{\mu}}\right)^{4}\\
\nu_{\mathrm{e1}}=r^{2}\Omega\max\left(\left(\frac{\Omega}{N_{\mathrm{T}}}\right)^{1/2}\left(\frac{\kappa}{r^{2}N_{\mathrm{T}}}\right)^{1/2},q^{2}\left(\frac{\Omega}{N_{\mathrm{T}}}\right)^{4}\right),\end{array}\label{eq:nu e0}\end{equation}
with $N_{\mu}$ and $N_{\mathrm{T}}$ respectively the {}``compositional''
and thermal part of the buoyancy frequency and $\kappa$ the opacity
(see \citealt{Spruit2002} for more details). Analogously, \citet{Spruit2002}
derive the following expression for the effective radial diffusivity
$D_{\mathrm{re}}$: \begin{equation}
\begin{array}{l}
D_{\mathrm{re}}=\frac{D_{\mathrm{e0}}D_{\mathrm{e1}}}{D_{\mathrm{e0}}+D_{\mathrm{e1}}}f(q),\\
D_{\mathrm{e0}}=r^{2}\Omega q^{4}\left(\frac{\Omega}{N_{\mu}}\right)^{6}\\
D_{\mathrm{e1}}=r^{2}\Omega\max\left(q\left(\frac{\Omega}{N_{\mathrm{T}}}\right)^{3/4}\left(\frac{\kappa}{r^{2}N_{\mathrm{T}}}\right)^{3/4},q^{4}\left(\frac{\Omega}{N_{\mathrm{T}}}\right)^{6}\right)\end{array}\label{eq:D re}\end{equation}
Analytic studies often use as approximation only $\nu_{\mathrm{e0}}$,
$D_{\mathrm{e0}}$ or $\nu_{\mathrm{e1}}$, $D_{\mathrm{e1}}$. Since
we study thermonuclear burning in a thin shell, both thermal and compositional
gradients are expected to be important in the shell. In our models
we, therefore, use expressions (\ref{eq:nu total}) and (\ref{eq:D re})
including both parts.

While the Tayler-Spruit dynamo is often applied in the spherical approximation,
we use it here for a thin shell (see also \citealt{Piro2007}). For
the dynamo to work, the presence of a non-zero shear \emph{locally}
is sufficient. Since the resulting magnetic field is predominantly
toroidal, the angular momentum transport and the chemical mixing induced
by the dynamo are limited to the radial region where the shear is
present.

In Section \ref{sub:Non-magnetic-mixing} we discuss some potential
problems with the Tayler-Spruit mechanism.

\subsection{Nuclear network}

An extensive nuclear network consisting of 35 isotopes and 65 nuclear
reactions is employed to model the chemical evolution. The most important
reaction sequence for our models is the triple alpha process, for
which the net reaction is: \begin{equation}
3\,^{4}\mathrm{He\rightarrow^{12}C}+\gamma.\label{eq:triple alpha}\end{equation}
Furthermore, alpha capture reactions onto carbon and oxygen are considered:
\begin{equation}
\begin{array}{l}
^{4}\mathrm{He}+^{12}\mathrm{C}\rightarrow^{16}\mathrm{O}+\gamma\\
^{4}\mathrm{He}+^{16}\mathrm{O}\rightarrow^{20}\mathrm{Ne}+\gamma.\end{array}\label{eq:alpha captures}\end{equation}
We disable the energy output of the $\mathrm{^{12}C+{}^{12}C}$ and
$\mathrm{^{12}C+{}^{16}O}$ reactions to prevent the occurrence of
unstable carbon burning at larger depths in our models. Any heating
that stable carbon may have provided can be modeled as an extra contribution
to the heat flux from the crust (see Section \ref{sub:Flux-from-the}).

\subsection{Viscous and compressional heating}

Viscous energy release due to the presence of shear heats the neutron
star envelope. The energy generation rate per unit mass by viscous
heating, $\epsilon_{\mathrm{visc}}$, can be calculated as \begin{equation}
\epsilon_{\mathrm{visc}}=\frac{\nu}{2}\left(q\Omega\right)^{2}.\label{eq:evisc}\end{equation}
 We find that the relatively small shear present in our models, in
the range of rotation rates and mass accretion rates we consider,
leads to a negligible amount of viscous heating, as the total energy
generated per unit time is several orders of magnitude smaller than
the assumed flux from the crust or the energy generated by stable
helium burning (see also \citealt{Piro2007}). Therefore, we do not
take viscous heating into account in our calculations.

The continuous accretion of matter onto the neutron star compresses
the matter in the envelope. This gives rise to compressional heating,
which is calculated in our models.

\subsection{Initial model}

We start with a grid of approximately 600 grid points that represent
the outer layers of the neutron star from a column depth of $y\simeq1\,\mathrm{g\, cm^{-2}}$
down to $y\simeq10^{11}\,\mathrm{g\, cm^{-2}}$, with a total mass
of $\simeq10^{24}\,\mathrm{g}$. As initial composition we take 50\%
$\mathrm{^{12}C}$ and 50\% $\mathrm{^{16}O}$. This acts as a buffer
which is inert with respect to the triple-alpha reactions of helium
burning. On this grid matter is accreted with a composition of 99\%
$\mathrm{^{4}He}$ and 1\% other isotopes, based on the equilibrium
mass fractions of the CNO cycle.

We use a core with a mass of $M=\mathrm{1.4\, M_{\odot}}$ and a radius
$R=10\,\mathrm{km}$ for a canonical neutron star. The angular velocity
of the core, $\Omega$, is expressed in units of the Keplerian angular
velocity at the surface: $\Omega_{\mathrm{K}}=1.4\cdot10^{4}\,\mathrm{radians\, s^{-1}}$.
We create models with different values for $\Omega$.

The heat flux from the neutron star crust is modeled by setting the
luminosity at the inner boundary of our grid to a fixed value $L_{\mathrm{crust}}$.

\subsection{Flux from the crust\label{sub:Flux-from-the}}

As the ignition of the accreted helium depends strongly on the temperature
and we do not consider heating by stable hydrogen burning, it is influenced
by $L_{\mathrm{crust}}$. The crust is a deeper layer of the neutron
star starting at $y\simeq10^{12}\,\mathrm{g\, cm^{-2}}$ and extending
down to the core where densities exceed the nuclear density. When
matter is accreted onto the neutron star, the crust is compressed.
This gives rise to pycnonuclear burning and electron-capture reactions
which heat both the neutron star core as well as the outer layers.
Therefore, this heat flux is generally assumed to scale with the accretion
rate and a fixed energy generated per accreted nucleon is applied.
Previous studies often use a value close to $0.1$\,MeV/nucleon (\citealt{Haensel2003}).
\citet{Cumming2006} use values in the range of 0.1--0.3\,MeV/nucleon
when studying the ignition conditions of long duration type I X-ray
bursts. The analysis of the superburst from 4U~1608-52 (\citealt{Keek2008})
indicates that the temperature of the crust was likely much higher
than predicted by current models. Recently \citet{Gupta2007} calculated
the thermal structure of the crust, for the first time including electron
captures into excited states, finding a heat flux from the crust up
to ten times higher than obtained in previous studies (see also \citealt{Gupta2008}).
It is clear that the value of the heat flux from the crust is still
not well determined. Therefore, we choose to vary it independently
from the mass accretion rate.

Heat from helium burning in the envelope may be conducted into the
core. From here it is either dissipated by neutrino emission or conducted
outwards, back to the envelope on a thermal time scale of the order
of $10^{4}\,\mathrm{yr}$. This gives an extra contribution to $L_{\mathrm{crust}}$,
the size of which depends on the amount of heat from the envelope
conducted inwards and on the efficiency of neutrino cooling of the
core. The latter is determined from the equation of state of the core
(see discussion in \citealt{Cumming2006}), which is not well constrained
at this moment (see, e.g., \citealt{Lattimer2007}). Therefore, a
certain value for $L_{\mathrm{crust}}$ can physically be obtained
by different combinations of an outward flux from the core and the
crust as well as an inward heat flux from the envelope. In our calculations
we do not go into these details, but treat $L_{\mathrm{crust}}$ as
a free parameter. When we make a comparison to the physical situation,
we will assume that crustal heating gives the domination contribution
to $L_{\mathrm{crust}}$.

By choosing a certain value for $L_{\mathrm{crust}}$, we fix the
total heat flow into and out of the envelope to a net flux into the
envelope. For studying unstable burning, a flash would give rise to
a large heat flow out of the envelope and our boundary condition would
not be valid for the duration of the flash. However, we consider only
stable burning, which leads to a relatively lower heat flux that is
constant in time and is compatible with our boundary condition.

\section{Results}

\subsection{Turbulent mixing}

The accretion of angular momentum gives rise to shear, which causes
turbulence and induces a magnetic field. These drive the transport
of angular momentum and the mixing of the chemical composition. In
turn, the chemical and angular velocity profiles influence the diffusivity
and viscosity, until an equilibrium is reached. We present the properties
of equilibrium models with different rotation rates of the core and
different mass accretion rates. The heat flux from the crust is set
to a value such that steady-state burning is achieved.

\begin{figure}
\includegraphics[clip,width=1\columnwidth]{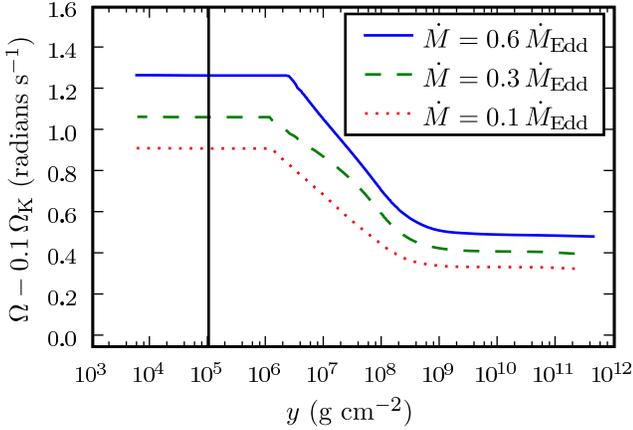}

\caption{\label{fig:Rotational-velocity-as}Rotational velocity $\Omega$ as
a function of column depth $y$ for three values of the mass accretion
rate and an angular velocity of the core of $0.1\,\Omega_{\mathrm{K}}$.
Accreted angular momentum is shared with the region to the left of
the vertical line.}

\end{figure}

Fig. \ref{fig:Rotational-velocity-as} shows the rotational velocity
profile for several values of the mass accretion rate. The viscous
processes transport angular momentum efficiently through the outer
layers of the neutron star into the core, resulting in a small difference
in $\Omega$ between the outer and inner part of our model of $\Delta\Omega/\Omega\sim10^{-4}$.
The largest change in $\Omega$ is at $y\simeq10^{8}\,\mathrm{g\, cm^{-2}}$.
We will see that at this depth helium burning generates heat and changes
the chemical composition. In the figure we indicate the region of
the model where the accreted angular momentum is added as described
in the previous section. Since the accreted angular momentum is quickly
shared with this region, the angular momentum transport is not calculated
in a physically correct way in that part of the model. Furthermore,
at $y\lesssim10^{7}\,\mathrm{g\, cm^{-2}}$ lies the pseudo-Lagrangian
region where the accreted matter is added. Therefore, we will limit
most plots to show the Lagrangian part, $y\gtrsim10^{7}\,\mathrm{g\, cm^{-2}}$.

\begin{figure}
\includegraphics[clip,width=1\columnwidth]{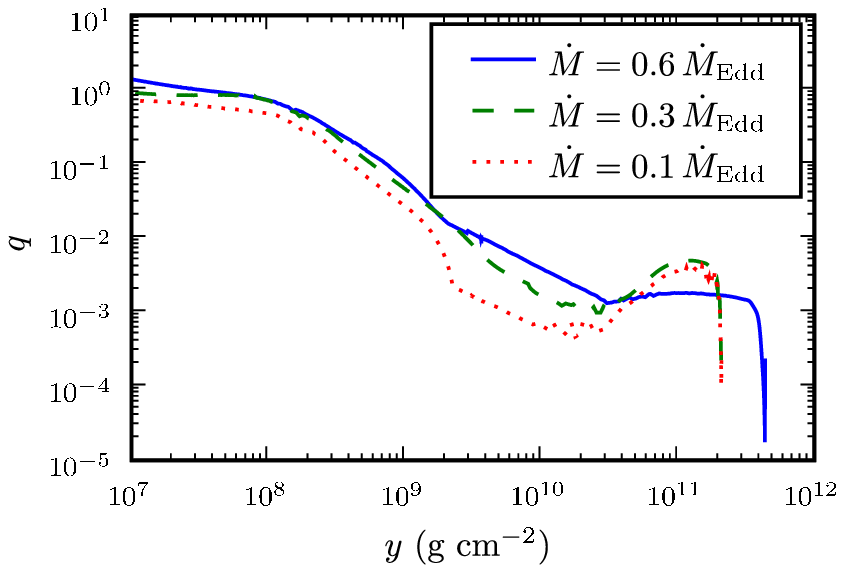}

\includegraphics[clip,width=1\columnwidth]{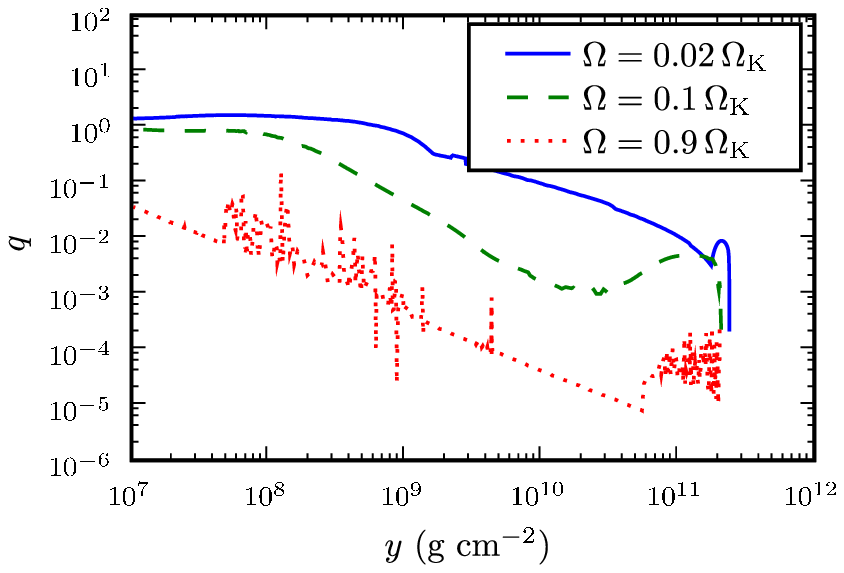}

\caption{\label{fig:Comparison-of-shear}Comparison of shear $q$ with column
depth $y$. \emph{Top}: models with different mass accretion rates
and their rotation rate of the core fixed to $\Omega=0.1\,\Omega_{\mathrm{K}}$.
Note that the model with $\dot{M}=0.6\,\dot{M}_{\mathrm{Edd}}$ accreted
for a longer time and, therefore, reaches larger depths. \emph{Bottom}:
models with different rotation rates of the core and a fixed accretion
rate of $\dot{M}=0.3\,\dot{M}_{\mathrm{Edd}}$. The model with $\Omega=0.9\,\Omega_{\mathrm{K}}$
suffers from numerical noise.}

\end{figure}

For higher accretion rates, angular momentum is added faster to the
star, resulting in a higher shear (Figure \ref{fig:Comparison-of-shear}
top), although the difference is less than an order of magnitude in
the range of accretion rates we consider. Since the accreted material
originally rotates with the Keplerian velocity, models with a more
slowly rotating core experience an up to $10^{3}$ times larger shear
(Fig. \ref{fig:Comparison-of-shear} bottom). The shear profiles show
two broad {}``bumps'': one in the region $10^{7}\lesssim y\lesssim10^{9}\,\mathrm{g\, cm^{-2}}$
and another at $y\gtrsim10^{10}\,\mathrm{g\, cm^{-2}}$. We will see
below that due to nuclear burning and diffusion there is a gradient
in the chemical composition in these regions. This gradient has a
stabilizing effect which reduces the angular momentum transport, causing
a somewhat larger shear. %

\begin{figure}
\includegraphics[bb=170bp 302bp 450bp 463bp,clip,width=1\columnwidth]{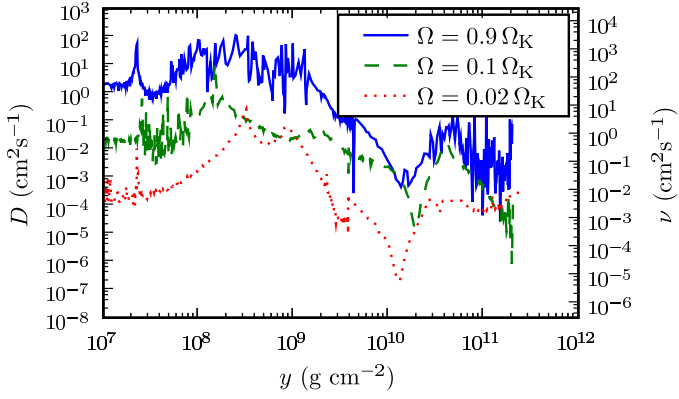}

\caption{\label{fig:Rotational-diffusivity-and}Non-magnetic diffusivity $D$
and viscosity $\nu$ due to Eddington-Sweet circulation as a function
of column depth $y$ for different values of the angular velocity
of the core $\Omega$ and an accretion rate of $\dot{M}=0.3\,\dot{M}_{\mathrm{Edd}}$.
Note that $\nu$ is larger than $D$ by a fixed factor of $30$. Especially
the model with $\Omega=0.9\,\Omega_{\mathrm{K}}$ suffers from numerical
noise.}

\end{figure}

\begin{figure}
\includegraphics[clip,width=1\columnwidth]{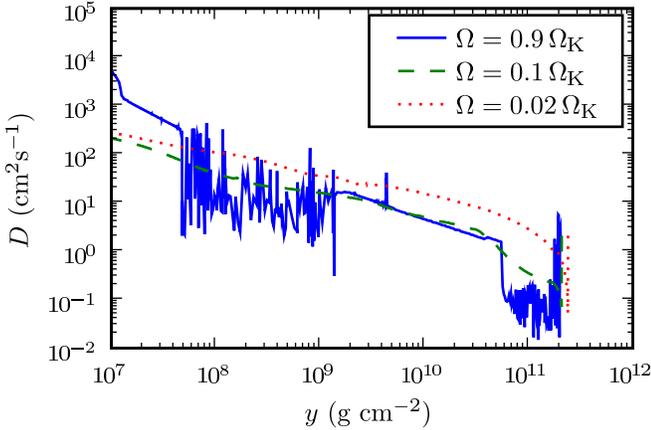}

\caption{\label{fig:Magnetic-diffusivity-as}Magnetic diffusivity $D$ as a
function of column depth $y$ for different values of the angular
velocity of the core and an accretion rate of $\dot{M}=0.3\,\dot{M}_{\mathrm{Edd}}$.
A minimum in $D$ is apparent at $\Omega\simeq0.1\,\Omega_{\mathrm{K}}$.
The model with $\Omega=0.9\,\Omega_{\mathrm{K}}$ suffers from numerical
noise.}

\end{figure}

\begin{figure}
\includegraphics[clip,width=1\columnwidth]{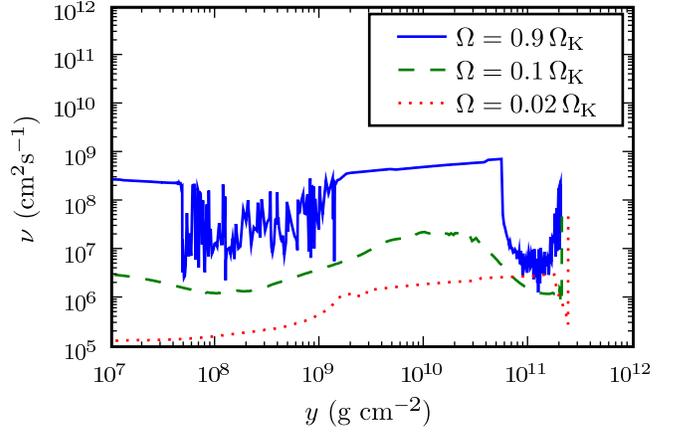}

\caption{\label{fig:Magnetic-viscosity-as}Magnetic viscosity $\nu$ as a function
of column depth $y$ for different values of the angular velocity
of the core $\Omega$. The model with $\Omega=0.9\,\Omega_{\mathrm{K}}$
suffers from numerical noise.}

\end{figure}

The presence of shear causes turbulence and induces a magnetic field,
that both drive diffusion and viscous processes. We obtain a magnetic
field with a radial component $B_{r}\simeq10^{5}\,\mathrm{G}$ and
a toroidal component $B_{\phi}\simeq10^{10}\,\mathrm{G}$. We calculate
$D$ and $\nu$ both due to the hydrodynamic instabilities and due
to the (generation of) the magnetic field (Fig. \ref{fig:Rotational-diffusivity-and},
\ref{fig:Magnetic-diffusivity-as} and \ref{fig:Magnetic-viscosity-as}).
We find the latter to have the largest contribution to the diffusivity
and viscosity. Of the non-magnetic instabilities, Eddington-Sweet
circulation is the dominant process. At $y\simeq10^{8}\,\mathrm{g\, cm^{-2}}$
$D$ and $\nu$ due to this process peak, since at this depth the
luminosity gradient is largest due to helium burning. Note that convective
mixing plays no role in our models of steady state burning, because
heat transport is radiative throughout our models, except at the onset
of a flash, similar to what is found by \citet{Woosley2004}.

We investigate how $D$ and $\nu$ depend on the rotation rate $\Omega$.
The contribution from the non-magnetic instabilities increases with
increasing $\Omega$ (Fig. \ref{fig:Rotational-diffusivity-and}).
For the Eddington-Sweet circulation the diffusivity and viscosity
are proportional to $\Omega^{2}$ (\citealt{Kippenhahn1974,Heger2000}).
 The diffusivity due to the magnetic field generation increases towards
lower rotation rates as, approximately, $D\propto\Omega^{-3/4}$ (Fig.
\ref{fig:Magnetic-diffusivity-as}).%
 The precise dependence is influenced by buoyancy effects from temperature
and compositional gradients (Sect. \ref{sub:Diffusivity-and-viscosity};
\citealt{Spruit2002}). The magnetic viscosity, which is induced by
the magnetic torques instead of MHD instabilities, increases towards
higher $\Omega$, since it depends mostly on the rotation rate: $D\propto\Omega^{3/2}$
(Fig. \ref{fig:Magnetic-viscosity-as}). Interestingly, while the
magnetic diffusivity dominates at most depths, for models with $\Omega\gtrsim0.1\Omega_{\mathrm{K}}$
Eddington-Sweet circulation gives a contribution to the diffusivity
which is equal or larger at the depth of $y\simeq10^{8}\,\mathrm{g\, cm^{-2}}$
where helium is burned.%

During our calculations we assume angular momentum is accreted with
an efficiency of 100\%. Due to viscous processes in the disk, the
actual efficiency may be lower. We investigate the dependency of the
shear, diffusivity and viscosity on the accretion efficiency by comparing
a model with 100\% to a model with 0\% efficiency. In the helium burning
region, the difference is at most several tens of percents. For more
realistic efficiencies the deviation from the 100\% efficient model
is likely even smaller. Therefore, we neglect this effect in our calculations.

\begin{figure}
\includegraphics[clip,width=1\columnwidth]{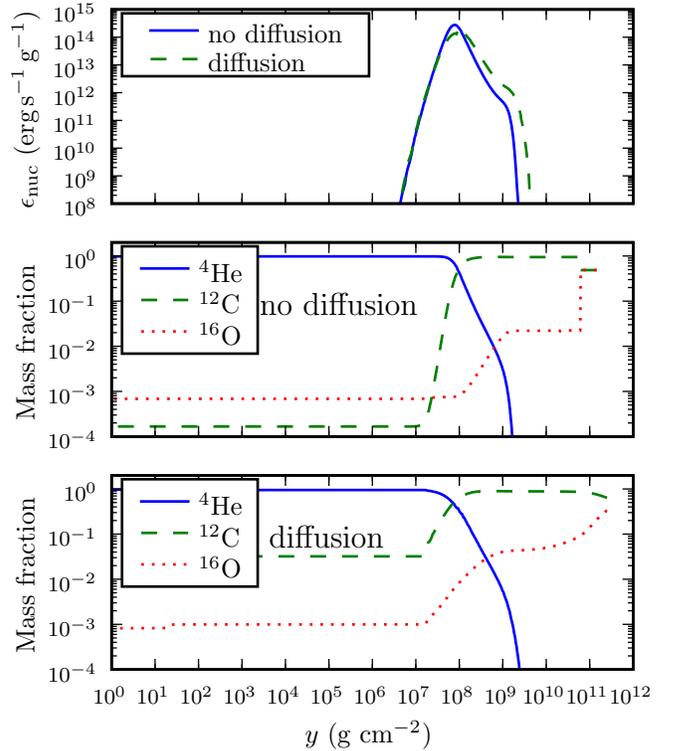}

\caption{\label{fig:Isotope-mass-fractions}Nuclear energy generation rate
$\epsilon_{\mathrm{nuc}}$ and isotope mass fractions as a function
of column depth $y$ for two models with $\dot{M}=0.3\,\dot{M}_{\mathrm{Edd}}$
and $\Omega=0.1\,\Omega_{\mathrm{K}}$. One model is calculated including
chemical diffusion, while the other is not.}

\end{figure}

\begin{figure}
\includegraphics[width=1\columnwidth]{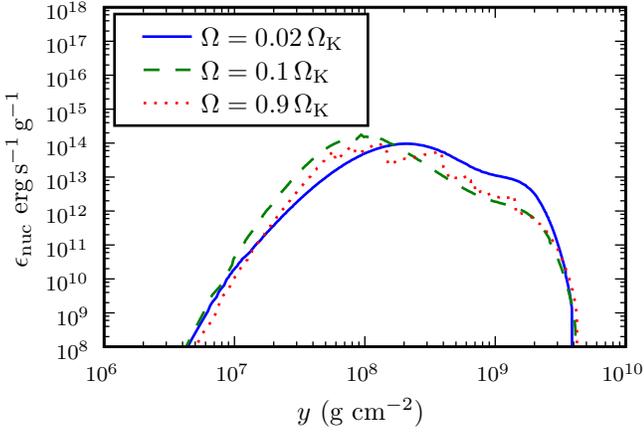}

\caption{\label{fig:Nuclear-energy-generation}Nuclear energy generation rate
$\epsilon_{\mathrm{nuc}}$ as a function of column depth $y$ for
different values of the angular velocity of the core $\Omega$. Models
with higher diffusivity (see Fig. \ref{fig:Magnetic-diffusivity-as})
have more He burning at higher column depths. The model with $\Omega=0.9\,\Omega_{\mathrm{K}}$
suffers from numerical noise.}

\end{figure}

The viscosity allows for the transport of angular momentum, leading
in the envelope to the angular velocity profile we presented in Fig.
\ref{fig:Rotational-velocity-as}. Similarly, diffusion mixes the
chemical composition of our model. Without diffusion, the chemical
profile consists on the outside of the star (Fig. \ref{fig:Isotope-mass-fractions}
middle, left-hand side) of accreted helium which burns at $y\simeq10^{8}\,\mathrm{g\, cm^{-2}}$,
forming a layer of mostly carbon. On the inside (right-hand side)
the composition is that of the initial model: 50\% $^{12}$C and 50\%
$^{16}$O. When we include diffusion, $^{4}$He is mixed down to higher
column depths , while some $^{12}$C is mixed all the way to the surface
(Fig. \ref{fig:Isotope-mass-fractions} bottom). Due to the increased
helium abundance at $y\gtrsim10^{8}\,\mathrm{g\, cm^{-2}}$, the energy
generation rate is higher there compared to the models without diffusion
(Fig. \ref{fig:Isotope-mass-fractions} top). For models with different
rotation rates of the core, we see that those with higher diffusivity
(see Fig. \ref{fig:Magnetic-diffusivity-as}) mix helium deeper inwards
and consequently release more energy due to thermonuclear burning
at higher column depths (Fig. \ref{fig:Nuclear-energy-generation}).

\begin{figure}
\includegraphics[clip,width=1\columnwidth]{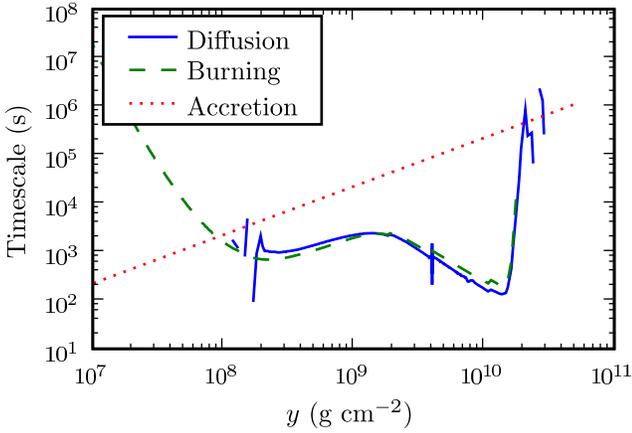}

\caption{\label{fig:Timescales-for-helium}Timescales for helium burning, chemical
diffusion and accretion as a function of column depth $y$ for $\Omega=0.1\,\Omega_{\mathrm{K}}$
and $\dot{M}=0.3\,\dot{M}_{\mathrm{Edd}}$.}

\end{figure}

In the steady state helium is burned typically at the depth where
the accretion timescale equals the burning timescale. In the case
of diffusive mixing of the chemical composition, fuel can be transported
to deeper layers than the accretion column. In this case, helium is
burned down to the depth where the burning timescale equals the diffusion
timescale. We determine as a function of column depth the burning
and chemical diffusion timescales $t_{\mathrm{nuc}}$ and $t_{\mathrm{diff}}$
for $^{4}$He by calculating \begin{equation}
t_{\mathrm{nuc}},\, t_{\mathrm{diff}}=Y\left(\frac{\Delta Y}{\Delta t}\right)^{-1},\label{eq:tnuc tdiff}\end{equation}
with $Y$ the $^{4}$He mass fraction and $\Delta Y$ the change in
$Y$ due to nuclear burning or chemical diffusion during one time
step $\Delta t$ of our calculations (Fig. \ref{fig:Timescales-for-helium}).
In the inner part of the burning zone, where helium is mixed in, the
chemical diffusion and burning timescales differ by at most a few
tens of percents, ensuring that steady-state burning can take place
down to larger depths than without diffusion. %

\subsection{Effect of mixing on stability}

The temperature in the burning shell has an important influence on
the stability of the nuclear burning. Typically, burning is stable
for higher temperatures of the neutron star envelope, which depends
highly on the heat flux from the crust. We express this heat flux
as a luminosity $L_{\mathrm{crust}}$ and investigate for which values
burning is stable by starting at a high value of $L_{\mathrm{crust}}$
and subsequently lowering it at a rate which is slow with respect
to the thermal timescale of the inner part of our model. For a certain
value of $L_{\mathrm{crust}}$ nuclear burning no longer proceeds
stably and a flash occurs. We refer to this as the critical luminosity
$L_{\mathrm{crit}}$. At that point we stop the calculation.

\begin{figure}
\includegraphics[bb=164bp 300bp 418bp 490bp,clip,width=1\columnwidth]{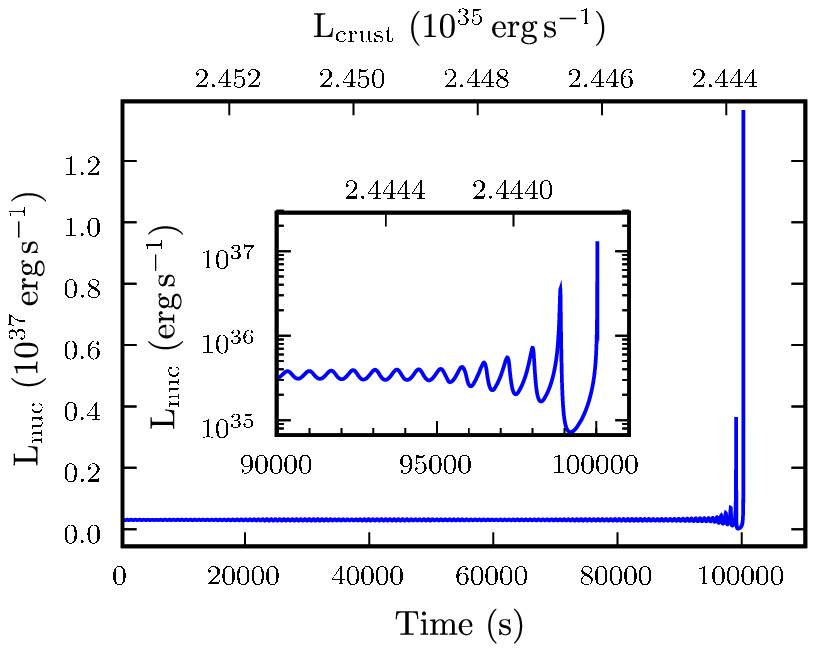}

\includegraphics[bb=164bp 300bp 418bp 490bp,clip,width=1\columnwidth]{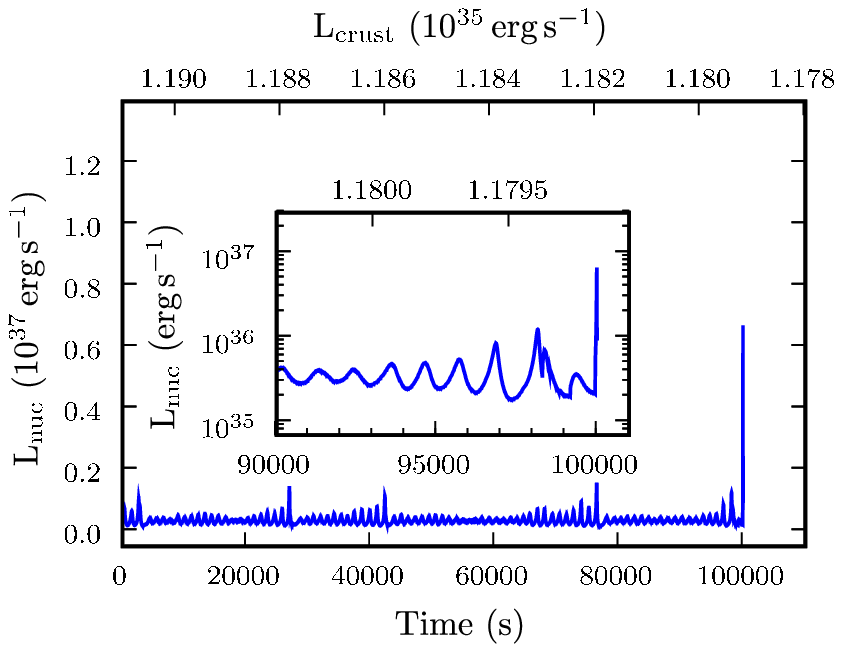}

\caption{\label{fig:Nuclear-luminosity-L}Nuclear luminosity $L_{\mathrm{nuc}}$
due to helium burning as a function of time, while lowering $L_{\mathrm{crust}}$
at a constant rate. Shown are the last $10^{5}\,\mathrm{s}$ before
the onset of unstable burning. The insets zoom in on the last $10^{4}\,\mathrm{s}$.
\emph{Top}: model without diffusion for $\Omega=0.1\,\Omega_{\mathrm{K}}$,
$\dot{M}=0.3\dot{M}_{\mathrm{Edd}}$. While burning is initially stable,
we see an increase in the amplitude of the oscillations, until finally
burning becomes unstable. \emph{Bottom}: similar model including diffusion.
The amplitude of oscillatory burning is much larger and less regular
than for the model without diffusion.}

\end{figure}

We perform this procedure for two models with $\Omega=0.1\,\Omega_{\mathrm{K}}$
and $\dot{M}=0.3\,\dot{M}_{\mathrm{Edd}}$. Both models take viscous
processes into account, such that the angular velocity profiles are
similar. However, we enable chemical diffusion only for one model,
while the other is calculated without chemical diffusion. For the
model excluding chemical mixing, we start with stable burning (Fig.
\ref{fig:Nuclear-luminosity-L} top).%
 Lowering $L_{\mathrm{crust}}$, helium shell burning exhibits oscillations
with increasing amplitude. These oscillations are well resolved in
time by our calculations. For $L_{\mathrm{crust}}=2.448\cdot10^{35}\,\mathrm{erg\, s^{-1}}$
the oscillation amplitude increases rapidly, until a flash occurs.
We regard this as the onset of unstable burning. It is possible that
what we regard as the onset of instability is actually just an oscillation
with a large amplitude, which could be followed by a series of oscillations
before the unstable situation is reached. Nevertheless, observing
how fast the amplitude of the oscillations increases in a small interval
of $L_{\mathrm{crust}}$, we expect that our measure of $L_{\mathrm{crit}}$
is accurate within $0.2$\%. The period of the pulses increases, from
approximately $6\cdot10^{2}\,\mathrm{s}$ at $L_{\mathrm{crust}}=2.448\cdot10^{35}\,\mathrm{erg\, s^{-1}}$
to $9\cdot10^{2}\,\mathrm{s}$ at $L_{\mathrm{crust}}=2.444\cdot10^{35}\,\mathrm{erg\, s^{-1}}$,
just before the final two oscillations.

For the model including diffusion, the behavior is similar (Fig. \ref{fig:Nuclear-luminosity-L}
bottom). The amplitude of the oscillations, however, is large for
a broader range of $L_{\mathrm{crust}}$ and the behavior is less
regular. During intervals of $10^{4}-10^{5}\,\mathrm{s}$ the amplitude
slowly increases and subsequently decreases either slowly or fast.
The cause of the irregularities is likely the increased complexity
of our model when turbulent mixing is included, giving rise to more
{}``noise''. Despite the irregular behavior, we can still discern
a general trend towards oscillations with larger amplitudes while
we lower $L_{\mathrm{crust}}$. The period of the oscillations is
$\sim1.2\cdot10^{3}\,\mathrm{s}$ and does not change significantly.
The value of $L_{\mathrm{crit}}$ we obtain for this model is over
a factor two smaller than for the model without diffusion: $L_{\mathrm{crit}}=1.180\cdot10^{35}\,\mathrm{erg\, s^{-1}}$.
This shows the substantial stabilizing effect of turbulent mixing.
The uncertainty in $L_{\mathrm{crit}}$ is likely larger than for
the model without diffusion due to the increased noise. Note that
for the chosen angular velocity of the core, $\Omega=0.1\,\Omega_{\mathrm{K}}$,
the diffusivity is at its minimum value. For another choice of $\Omega$,
the effect is likely to be even larger.

\begin{figure}
\includegraphics[bb=175bp 302bp 450bp 464bp,clip,width=1\columnwidth]{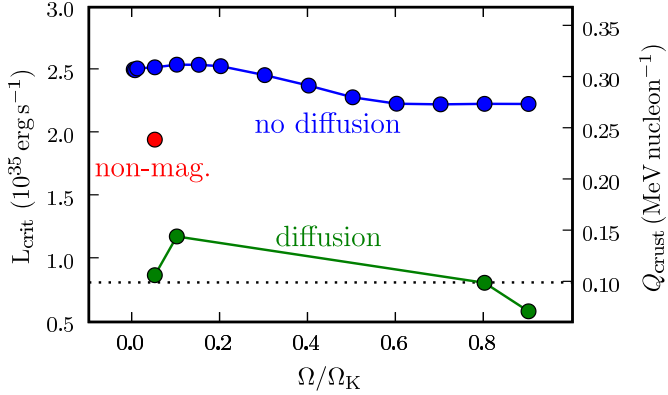}

\caption{\label{fig:The-heat-flux}The heat flux from the crust at the transition
of stable to unstable burning $L_{\mathrm{crit}}$ as a function of
the angular velocity $\Omega$ of the core in units of the Keplerian
velocity at the surface for models with $\dot{M}=0.3\dot{M}_{\mathrm{Edd}}$.
Each dot represents one model. Two series of models are calculated:
without (upper) and with chemical diffusion (lower). One model is
calculated without the effect of magnetic fields (middle). Here the
secular shear instability drives the chemical mixing. On the right-hand
vertical axis is indicated the heat generated in the crust per accreted
nucleon, $Q_{\mathrm{crust}}$, for the given accretion rate. The
dotted line indicates $Q_{\mathrm{crust}}=0.1\,\mathrm{MeV\, nucleon^{-1}}$,
which is used in many studies.}

\end{figure}

We investigate the dependence of $L_{\mathrm{crit}}$ on $\Omega$
by calculating two series of models, including and excluding chemical
diffusion. Due to the increased noise, the models which include diffusion
are computationally more expensive. Therefore, we limit ourselves
to two models with relatively low rotation rates, $\Omega=0.05\Omega_{\mathrm{K}},\,0.1\Omega_{\mathrm{K}}$,
and two models with high rotation rates: $\Omega=0.8\Omega_{\mathrm{K}},\,0.9\Omega_{\mathrm{K}}$.
We find that $L_{\mathrm{crit}}$ is lower by more than a factor of
$2$ for all values of $\Omega$ we consider (Fig. \ref{fig:The-heat-flux}).
Furthermore, even for the models without diffusion we see a decrease
in $L_{\mathrm{crit}}$ of up to $12\%$ when going to higher rotation
rates. This stabilizing effect is due to the centrifugal force $F_{\mathrm{cf}}\propto\Omega^{2}$,
which lowers the effective gravity, increasing the efficiency of stabilizing
effects. Note that due to the one-dimensional approximation we use,
this is likely an underestimation (see Sect. \ref{sub:Centrifugal-force}). 

We noted earlier that diffusivity due to Eddington-Sweet circulation
can be comparable to or even exceed the diffusivity due to magnetic
fields. To investigate the effect of non-magnetic instabilities alone,
we created a model with $\Omega=0.05\Omega_{\mathrm{K}}$ and $\dot{M}=0.3\dot{M}_{\mathrm{Edd}}$
including only angular momentum transport and chemical diffusion due
to non-magnetic instabilities. Angular momentum transport is much
less efficient, leading to a large shear. The secular shear instability
drives the diffusivity and viscosity. For this model we find $L_{\mathrm{crit}}=1.948\cdot10^{35}\,\mathrm{erg\, s^{-1}}$,
which is $23\%$ lower than what we find without diffusion (Fig. \ref{fig:The-heat-flux}).

When we change $L_{\mathrm{crust}}$ it takes a thermal timescale
for our model to adjust. Since we lower $L_{\mathrm{crust}}$ continuously,
this introduces a systematic error in the value of $L_{\mathrm{crit}}$
we obtain, as the model is not fully adjusted to that value of $L_{\mathrm{crust}}$
when we stop the calculation. To estimate the systematical error we
take one model and lower $L_{\mathrm{crust}}$ at different rates,
all slow with respect to the thermal timescale of the model, and find
that the difference in $L_{\mathrm{crit}}$ and, ergo, the systematic
error is about $4\%$.

\subsection{Mass accretion rate dependence}

\begin{figure}
\includegraphics[clip]{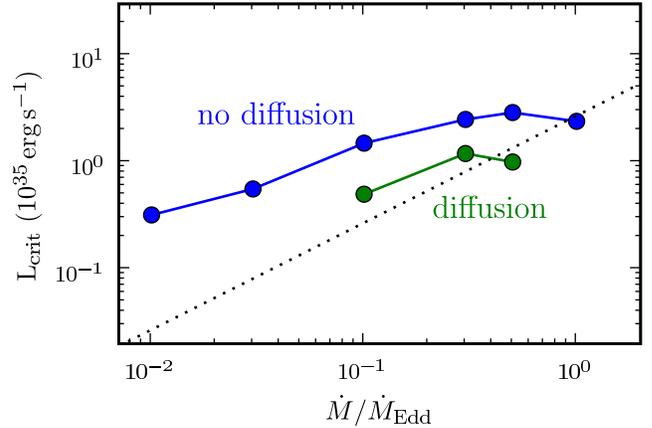}

\caption{\label{fig:The-heat-flux-mdot}The heat flux from the crust at the
transition of stable to unstable burning $L_{\mathrm{crit}}$ as a
function of the accretion rate $\dot{M}$ for models with $\Omega=0.1\,\Omega_{\mathrm{K}}$.
Each dot represents one model. Two series of models are calculated:
without (upper) and with chemical diffusion (lower). The dotted line
indicates the luminosity from the crust as a function of $\dot{M}$
for an energy release of $Q_{\mathrm{crust}}=0.1\,\mathrm{MeV\, nucleon^{-1}}$,
which is used in many studies.}

\end{figure}

The dependence of $L_{\mathrm{crit}}$ on the mass accretion rate
is investigated by calculating two series of models with and without
chemical diffusion, in a similar way as we studied the dependence
on $\Omega$. The angular velocity of the core is set to $0.1\Omega_{\mathrm{K}}$.
We find that $L_{\mathrm{crit}}$ increases towards higher mass accretion
rates (Fig. \ref{fig:The-heat-flux-mdot}). For the models without
chemical diffusion, the accretion rate above which the luminosity
of the crust exceeds $L_{\mathrm{crit}}$ as expected from a crustal
energy release of $Q_{\mathrm{crust}}=0.1\,\mathrm{MeV\, nucleon^{-1}}$,
which is used in many studies (see Sect. \ref{sub:Flux-from-the}),
lies at $\dot{M}_{\mathrm{Edd}}$, which is lower than found in previous
studies (e.g., \citealt{Bildsten1998}). Including chemical diffusion,
the transition takes place at lower rate: approximately $0.4\,\dot{M}_{\mathrm{Edd}}$.

To understand the behavior of increasing $L_{\mathrm{crit}}$ for
larger values of $\dot{M}$, we follow the stability analysis of thermonuclear
burning in a thin shell by \citet{Giannone1967} (see also \citealt{Yoon2004a}).
These authors derive an expression for the inverse of the growth time
scale of temperature perturbations, $\tau_{\mathrm{pert}}^{-1}$,
by considering both thermonuclear burning and radiative cooling (equation
16 in \citealt{Giannone1967}). In this expression the luminosity
from below the thin shell is ignored. We can include $L_{\mathrm{crust}}$
by adding it to the luminosity of the shell, which results in an extra
term in the numerator of the right-hand side of equation (16) in \citet{Giannone1967}:
$\tau_{\mathrm{pert}}^{-1}\propto-L_{\mathrm{crust}}/\Delta M$. Here,
$\Delta M$ is the mass of the shell, which for steady-state burning
is proportional to the mass accretion rate. Increasing $\dot{M}$
leads to a lower perturbation growth time scale. Therefore, to keep
the same perturbation growth time scale, a larger value of $L_{\mathrm{crust}}$
is required at larger $\dot{M}$. This simple picture suggests that
$L_{\mathrm{crit}}$ depends linearly on $\dot{M}$, while Fig. \ref{fig:The-heat-flux-mdot}
suggests for the series of models without chemical diffusion that
the relation deviates from a linear dependence. In fact, for larger
$\dot{M}$ also the temperature and density in the shell are larger,
which influences the perturbation growth time scale as well.

\section{Discussion}

We performed detailed calculations of thin shell burning of accreted
helium on a neutron star. Similar to what was found by \citet{Yoon2004}
for white dwarfs, also for neutron stars rotational mixing leads to
more stable burning. As was demonstrated by \citet{Piro2007}, rotationally
induced magnetic fields provide in most cases the dominant contribution
to the diffusivity and viscosity. \citet{Yoon2004} considered only
non-magnetic hydrodynamic instabilities. In those models the transport
of angular momentum is less efficient, resulting in a much larger
shear. The secular shear instability is in that case the driver of
turbulent mixing. We take magnetic fields into account through the
Tayler-Spruit dynamo (\citealt{Spruit2002}), which causes efficient
angular momentum transport and turbulent mixing. Eddington-Sweet circulation
also plays an important role, especially in the burning zone, at the
peak of the energy generation rate, where there is a steep gradient
in the luminosity.

In our models with diffusion the geometrical thickness of the burning
shell is larger (Fig. \ref{fig:Isotope-mass-fractions}). For instance,
for a model with $\Omega=0.1\,\Omega_{\mathrm{K}}$ and $\dot{M}=0.3\dot{M}_{\mathrm{Edd}}$
the thickness of the burning shell is $6.7\,\mathrm{m}$ excluding
and $8.7\,\mathrm{m}$ including chemical diffusion, where we define
the shell such that at the boundary the energy generation rate is
$10^{-3}$ times the peak value. This results in an increased stability
of thermonuclear burning.

Due to the increased stability, the critical value of the heat flux
for the crust $L_{\mathrm{crit}}$ at the transition of stable and
unstable burning is lower for models that include chemical diffusion.
Even for the rotation rate which gives the minimum diffusivity, nuclear
burning proceeds stably down to values of $L_{\mathrm{crit}}$ that
are a factor of two lower than for the non-diffusion case. In these
calculations we vary the flux from the crust independently from the
mass accretion rate. For several values of $\dot{M}$ and $\Omega$
we determine $L_{\mathrm{crit}}$. Observationally, this boundary
is placed at $\dot{M}\simeq0.1\dot{M}_{\mathrm{Edd}}$. At this accretion
rate, we find $L_{\mathrm{crit}}=0.508\cdot10^{35}\,\mathrm{erg\, s^{-1}}$
for a model with $\Omega=0.1\Omega_{\mathrm{K}}$. Assuming the heat
flux from the crust is the main source of this luminosity, this corresponds
to an emergent heat from the crust of $0.27\,\mathrm{MeV\, nucleon^{-1}}$.
This is almost three times the value of $0.1\,\mathrm{MeV\, nucleon^{-1}}$
used in many previous studies. Nevertheless, it is in agreement with
the increased heat deposition in the crust from calculations by \citet{Gupta2007}. 

It is especially interesting to note that we obtain this result already
at the value for $\Omega$ for which the diffusivity is minimal. \citet{Piro2007}
find this behavior only at low rotation rates: $\Omega\lesssim0.1\Omega_{\mathrm{K}}$.
The Taylor-Spruit mechanism considers stabilizing effects due to temperature
and compositional stratification. \citet{Piro2007} simplify their
models by taking into account only the former when determining the
recurrence time of bursts and the condition for stable burning. This
allows for a simple dependence of $D$ on $\Omega$ of $D\propto\Omega^{-3/4}$.
However, by including both stabilizing effects the dependence of $D$
on $\Omega$ and $q$ is more complex. In our calculations we find
that the spread in the diffusivity in the considered range of rotation
rates is at most one order of magnitude. Furthermore, we find that
at high angular velocities Eddington-Sweet circulations drive mixing
just as efficiently as the magnetic instabilities at low rotation
rates. This means that for all rotation rates the stability of shell
burning is increased by chemical mixing due to rotation and rotationally
induced magnetic fields.

\subsection{Mixing efficiencies\label{sub:Non-magnetic-mixing}}

In our models we find that the magnetic instabilities and torques
provide the most efficient mixing of the chemical composition and
angular momentum. The only exception is found at higher rotation rates,
where the Eddington-Sweet circulations provide an important contribution
to the diffusivity in the helium burning shell. Note that we employ
a one-dimensional approximation of the diffusivity due to Eddington-Sweet
circulations (\citealt{Kippenhahn1974}). At higher rotation rates
two-dimensional effects may be important that cannot be tested with
our model, but that may influence the efficiency of mixing due to
these circulations.

Several studies need the Tayler-Spruit dynamo to reproduce observed
quantities: \citet{Heger2005a} model the pre-supernova evolution
of massive stars and use the angular momentum transport due to rotationally
induced magnetic fields to explain observed pulsar frequencies. \citet{Suijs2008}
performed similar calculations for white dwarf progenitors and find
that angular momentum transport by magnetic torques is needed to explain
the observed range of white dwarf rotation rates. 

However, the mechanism of the Tayler-Spruit dynamo is uncertain. \citet{Zahn2007}
performed calculations with three-dimensional models which do not
reproduce the dynamo effect. Therefore, it is possible that this mechanism
has a lower efficiency or is absent (see, e.g., the discussion in
\citealt{Maeder2008}). In that case, non-magnetic instabilities drive
chemical mixing and angular momentum transport. These are less efficient
by many orders of magnitude. This leads to a large shear, causing
the secular shear instability to provide the dominant contribution
to $D$ and $\nu$, similarly to what \citet{Yoon2004} found in white
dwarf models which did not include a rotationally induced magnetic
field. We performed calculations on one model without a rotationally
induced magnetic field and we find a value of $L_{\mathrm{crit}}$
that is $23\%$ lower than for similar models that do not take chemical
diffusion into account (Fig. \ref{fig:The-heat-flux}). This model
has a rotation rate of $0.05\Omega_{\mathrm{K}}$. Models with a slower
rotating core will have a higher shear, which may result in a lower
value of $L_{\mathrm{crit}}$. Therefore, even without magnetic fields,
the effect of rotational mixing on the stability of helium shell burning
is considerable.

\subsection{Centrifugal force\label{sub:Centrifugal-force}}

Another stabilizing effect of rotation is provided by the centrifugal
force. \citet{Yoon2004} found that for shell burning in white dwarfs,
the centrifugal force lowers the density and degeneracy in the shell,
which leads to increased stability. We created a series of models
without chemical mixing and found that for increasing $\Omega$, $L_{\mathrm{crit}}$
is lower by up to $12\%$. However, this likely is an underestimation.
Our implementation of the centrifugal force is a one-dimensional approximation.
The centrifugal force depends on latitude, being largest at the equator
and vanishing at the poles. Our calculations use a value which is
averaged over latitude (e.g. \citealt{Heger2000}). In the case of
rigidly rotating polytropes this approximation has been found accurate
for rotation rates up to $0.6\Omega_{\mathrm{K}}$ (\citealt{Yoon2004}
and references therein). Therefore, we limit the effect of the centrifugal
force above $60$\% of the critical velocity. In actuality, the centrifugal
force may reduce $L_{\mathrm{crit}}$ even further for fast spinning
neutron stars.

\subsection{Marginally stable burning}

\citet{Revnivtsev2001} observe from the LMXBs 4U~1636--53, 4U~1608--52
and Aql~X-1 millihertz quasi-periodic oscillations (mHz QPOs) and
speculate that this phenomenon may be related to a special mode of
nuclear burning. This behavior is thought to occur at the boundary
of stable and unstable burning, since it is observed in a narrow range
of accretion rates, where at lower accretion rates type I X-ray bursts
were observed while they were absent at higher accretion rates. The
oscillatory burning mode at this boundary was expected from one-zone
models by \citet{Paczynski1983} as {}``marginally stable burning''.
\citet{Heger2005} reproduce this behavior in multi-zone numerical
models of the neutron star envelope and derive the period of the oscillations
$P_{\mathrm{osc}}$ to be the geometric mean of the accretion and
thermal timescales: $P_{\mathrm{osc}}=\sqrt{t_{\mathrm{acc}}t_{\mathrm{therm}}}$. 

\citet{Altamirano2008} observe frequency drifts in mHz QPOs from
4U~1636--53. The mHz QPOs are observable for approximately $10^{4}\mathrm{s}$.
During this time the frequency drops by several tens of percents.
At a certain point a type I X-ray burst is observed, after which no
oscillations are detected for some time. The change in period is not
found to be directly related to variations of the X-ray flux on short
timescales. Therefore the frequency of the oscillations does not appear
to be sensitive to variations in the accretion rate. Furthermore,
\citet{Molkov2005} observed a long type I X-ray burst from SLX~1735--269
which exhibits $\sim10\,\mathrm{mHz}$ oscillations during the burst
decay. Also in this case a drift to lower frequencies takes place
during the part of the X-ray light curve where the black body temperature
is observed to decrease.

Also in our models we observe oscillations at the boundary of stable
and unstable burning. This is most pronounced in the models without
chemical diffusion. When we lower $L_{\mathrm{crust}}$, effectively
cooling our burning layer, the frequency of the oscillations decreases
by several tens of percents until a flash occurs. The frequency drift
takes place over a small range of $L_{\mathrm{crust}}$, just before
the onset of a flash. The amplitude of the oscillations is larger
in this range, making the oscillations more easily detectable. This
supports the suggestion by \citet{Altamirano2008} that the frequency
drifts are due to cooling of deeper layers and could explain why it
is not sensitive to short-term variations of the accretion rate. The
cooling could, for instance, be caused by the slow release of energy
from a deeper layer that was heated up during an X-ray burst.

Note that the models that include diffusion display oscillations as
well, but the behavior is much more irregular, as the amplitude of
the oscillations changes in a seemingly erratic way. This makes it
difficult to draw conclusions from the oscillatory behavior. We suspect
that numerical noise is the source of this behavior. A different implementation
of the mixing processes in the model may be required to improve this.

\subsection{Further consequences of mixing}

Apart from the increased stability of helium burning, chemical mixing
can give rise to other effects. Superbursts, which are energetic type
I X-ray bursts that last up to a day, are thought to be the result
of a carbon flash (\citealt{Strohmayer2002,Cumming2001}). Fifteen
superbursts have been observed from ten sources, most of which have
accretion rates $\dot{M}\gtrsim0.1\dot{M}_{\mathrm{Edd}}$ (\citealt{Kuulkers2003a,intZand2004,rem05,Kuulkers2005}).
As noted by \citet{Piro2007}, turbulent mixing could allow for stable
burning of helium at these accretion rates, building up a thick carbon
layer. However, all superbursters exhibit normal type I bursts as
well. In the presence of hydrogen, the rp-process creates heavy isotopes,
while lowering the carbon abundance. It has proven difficult to explain
how carbon can survive many normal bursts to fuel a superburst (e.g.,
\citealt{Keek2008}).

While helium is mixed down, we find that carbon is mixed up to smaller
depths (Fig. \ref{fig:Isotope-mass-fractions}). For our helium-accreting
model this is of little consequence, but many neutron stars in LMXBs
accrete hydrogen-rich material. Hydrogen burns via the hot-CNO cycle.
The energy generation rate of this process depends on the mass fractions
of the CNO seed nuclei, which are increased if carbon is mixed into
the hydrogen/helium layer. This may lower the hydrogen mass fraction
in the burning shell. If subsequently a hydrogen/helium flash occurs,
there is less hydrogen present for the rp-process to destroy the carbon
created by helium fusion. This could allow for the creation of a thick
carbon layer to fuel a superburst. Further calculations are needed
to assess the magnitude of this effect.

Our models show that carbon is mixed all the way to the neutron star
photosphere at $y=\mathrm{1\, g\, cm^{-2}}$. In case hydrogen-rich
material is accreted, the ashes of hydrogen and helium burning can
contain heavy isotopes, e.g. iron. If mixing increases the mass fractions
of these isotopes near the surface, this would allow for the detection
of absorption lines from the neutron star. Measuring the gravitational
redshift of the absorption line provides constraints on the neutron
star equation of state, such as has been shown with the tentative
result of \citet{Cottam2002} (\citealt{Ozel2006}). This is analogous
to the calculations of \citet{Weinberg2006}, who predict that convection
during particularly luminous X-ray bursts can mix heavy elements to
the photosphere.

\section{Conclusion}

A long-standing problem for the theory of thermonuclear burning in
the envelope of accreting neutron stars is the accretion rate at which
the burning changes from unstable to stable. While observations point
to $\dot{M}\simeq0.1\dot{M}_{\mathrm{Edd}}$, the models find $\dot{M}\simeq\dot{M}_{\mathrm{Edd}}$
when hydrogen-rich material is accreted and $\dot{M}\simeq10\dot{M}_{\mathrm{Edd}}$
when it is hydrogen-deficient. Recent analytical and numerical models
(\citealt{Piro2007}) suggest that turbulent mixing due to rotationally
induced magnetic fields leads to an increased stability of the thin
shell burning on neutron stars, similar to the effect of mixing due
to hydrodynamic instabilities in the white dwarf case (\citealt{Yoon2004}). 

We perform for the first time detailed calculations of nuclear burning
of accreted helium in the neutron star envelope using an extensive
numerical model including both magnetic and non-magnetic hydrodynamic
instabilities. In most cases the magnetic instabilities and torques
provide the largest contribution to the diffusivity and the viscosity,
respectively, while Eddington-Sweet circulations become important
at high rotation rates. We create models with an accretion rate of
$\dot{M}=0.3\dot{M}_{\mathrm{Edd}}$ and different rotation rates
of the core. We do not couple the heat flux from the crust to the
mass accretion rate, but leave it as a free parameter, finding the
threshold value between stable and unstable burning. This value is
lower by at least a factor of two for models including diffusion,
in comparison with models without chemical mixing. If we ignore the
rotationally induced magnetic fields, mixing due to the secular shear
instability can lower the critical heat flux by several tens of percents.
For larger rotation rates the centrifugal force allows for stable
burning at crustal heat fluxes that are up to 12\% smaller in our
models. Note, however, that this is likely an underestimation, because
of the employed approximation of the centrifugal force in our one-dimensional
model.

These effects show that the stability of helium shell burning is increased
due to rotation, such that stable burning can take place at sub-Eddington
accretion rates. Magnetic field generation through the Tayler-Spruit
dynamo yields the strongest effect. However, even without magnetic
fields, hydrodynamic instabilities and the centrifugal force increase
stability substantially. We calculate models including chemical diffusion
for several values of the accretion rate and find that at approximately
40\% of the Eddington-limited rate the critical heat flux from the
crust equals the value inferred from an energy release of $Q_{\mathrm{crust}}=0.1\,\mathrm{MeV\, nucleon^{-1}}$,
which is used in many previous studies. Therefore, to explain the
observed transition at $0.1\,\dot{M}_{\mathrm{Edd}}$ an increased
energy release is required, as found recently by \citet{Gupta2007}.
Combined with this larger energy release, rotational mixing and rotationally
induced magnetic fields may explain the transition from unstable to
stable burning at the observed accretion rate. A better understanding
of the mixing efficiencies is required to further constrain the importance
of these mixing processes.

Our models exhibit oscillations close to the transition of stable
to unstable burning, which have been related to the observed mHz QPOs
(\citealt{Heger2005}). When we lower the heat flux from the crust
at a constant rate, we find that the frequency of the oscillations
is reduced by up to several tens of percents. This supports the suggestion
by \citet{Altamirano2008} that the frequency drifts observed in mHz
QPOs are due to cooling.
\begin{acknowledgements}
LK acknowledges support from the Netherlands Organization for Scientific
Research (NWO). We thank D. Altamirano, P. G. Jonker and A. Heger
for interesting discussions and we are grateful for helpful comments
from the referee.
\end{acknowledgements}
\bibliographystyle{aa}
\bibliography{paper2}

\end{document}